# About conservation the angular momentum for asymmetric tensors in electrodynamics

Yurii A. Spirichev

The State Atomic Energy Corporation ROSATOM, "Research and Design Institute of Radio-Electronic Engineering" - branch of Federal Scientific-Production Center "Production Association "Start" named after Michael V.Protsenko",

Zarechny, Penza region, Russia

E-mail: yurii.spirichev@mail.ru

(Dated: August 6, 2017)

**Abstract**

It is customary to assume that the law of conservation of the angular momentum is violated for an asymmetric energy-momentum tensors. This is the reason for criticizing the Minkowski tensor and other asymmetric energy-momentum tensors. In this paper, it is shown that the laws of conservation of energy and momentum following from an asymmetric tensor in the form of its total divergence are equivalent to the divergence of its symmetric part. It is shown that the total divergence of the antisymmetric part of the asymmetric tensor is identically zero. From this, it follows that for the asymmetric energy-momentum tensor the law of conservation of the angular momentum is also fulfilled. It is shown that the linear invariant of the Minkowski tensor for a vacuum does not correspond to the quadratic invariant of the electromagnetic field. This indicates that the linear invariant of the Minkowski tensor and the three-dimensional stress tensor are not correct.

**1. Introduction**

The tensors of energy-momentum (EMT) play a fundamental role in the description of nature. Of them, follow the equations of laws of conservation of energy and momentum. The problem of choosing the EMT and the shape of the electromagnetic momentum of the interaction of the electromagnetic field with the medium has a long history and is the subject of numerous research and discussions. In recent years on this issue published papers [1-51]. In these articles discusses the forms EMT, the electromagnetic momentums in the medium, the power of Abraham and other electromagnetic forces. In discussions on this issue, typically consider Minkowski and Abraham EMT. The main argument against Minkowski EMT is its asymmetry. The asymmetrical EMT is credited with the violation of the law of conservation of angular momentum. This negative feature makes researchers, starting with Abraham, to build different variants of the symmetric EMT [52-57]. Electromagnetic momentum is a part of EMT, so his form is also the subject of debate. In these discussions competing forms of Minkowski momentums and the Abraham. The density of the electromagnetic momentum of Minkowski has the form $\mathbf{g}^M = \mathbf{D} \times \mathbf{B}$. The density of the electromagnetic momentum of Abraham has the form $\mathbf{g}^A = \mathbf{S}/c^2 = \mathbf{E} \times \mathbf{H}/c^2$. Here $\mathbf{E}$ and $\mathbf{D}$ - respectively, the electric field strength and electric induction; $\mathbf{H}$ and $\mathbf{B}$ are, respectively, the magnetic field intensity and magnetic induction. There are a number of experimental work for determining which of these form conform better to the reality. In paper [1] reviewed the experiments work on this problem and made their analysis, from which it follows that hitherto not obtained reliable experimental evidence of the correctness of any one of these forms of electromagnetic momentum. Thus, until recently, this fundamental question of electrodynamics remained open. In paper [4] from the tensors of the electromagnetic field and electromagnetic induction mathematically strictly obtained the EMT of the interaction of electromagnetic field with the dielectric medium, from which it follows the equations of conservation and the wave equations for energy and momentum. In paper [5] obtained the EMT of the interaction of electromagnetic field with a conducting environment (charges and currents). These EMT are asymmetric, leading to criticism from supporters of the use of symmetric tensors. Indeed, from a symmetric tensor of mechanical energy-momentum follows be covariant equation of conservation of mechanical momentum and angular momentum $\partial_t \mathbf{p}^M + \partial_m \sigma_{mn}^M = \mathbf{F}_{ext}^E$, where $\sigma_{mn}^M$ - a three-dimensional tensor of flux density of mechanical momentum, or stress tensor; $\mathbf{F}_{ext}^E$ - is the electric force, which are third-party forces. The covariant equation of conservation of electromagnetic momentum $\mathbf{p}^E$ in a continuous medium has the form [58, 59] $\partial_t \mathbf{p}^E + \partial_m \sigma_{mn}^E = \mathbf{F}_{ext}^M$, where $\sigma_{mn}^E$ - a three-dimensional tensor of flux density of electromagnetic momentum or stress tensor; $\mathbf{F}_{ext}^M$ - the mechanical forces that are third-party forces. In this case, mechanical forces are a reaction of the environment on electromagnetic forces. We can write make a complete covariant equation of balance of mechanical and electromagnetic forces $\mathbf{F}_{ext}^M = \mathbf{F}_{ext}^E$ or $\partial_t \mathbf{p}^E + \partial_m \sigma_{mn}^E = \partial_t \mathbf{p}^M + \partial_m \sigma_{imn}^M$. More generally this equation can be written in the form of a four-dimensional divergence of the tensor of energy-momentum for each of the indices [6]: $\partial^\mu T_{\mu\nu}^E = \partial^\mu T_{\mu\nu}^M$ and $\partial^\nu T_{\mu\nu}^E = \partial^\nu T_{\mu\nu}^M$, where $T_{\mu\nu}^E$ and $T_{\mu\nu}^M$, respectively, electromagnetic and mechanical the energy-momentums tensors. If $T_{\mu\nu}^E$ is a symmetric tensor, then there are no questions, since this equation is the equality of divergences of symmetric tensors. However, if it $T_{\mu\nu}^E$ is asymmetric tensor, this question requires special consideration. The purpose of this article is to examine the issue of conservation of electromagnetic energy, momentum and angular momentum for asymmetric EMTs.

## 2. The asymmetric tensors of electromagnetic energy-momentum

The canonical EMT written in the General form:

$$T_{\mu\nu} = \begin{bmatrix} W & i\frac{1}{c}\mathbf{S} \\ ic\cdot\mathbf{g} & t_{ik} \end{bmatrix} \quad (\mu, \nu = 0, 1, 2, 3; \; i, k = 1, 2, 3) \quad (1)$$

where $W$ – energy density;

$\mathbf{S}$ – Energy flux density (Poynting vector);

$\mathbf{g}$ – Momentum density;

$t_{ik}$ – tensor of the momentum flux density (the stress tensor)

The second rank tensor has a divergence for each of the indexes:

$$\partial^{\mu}T^{E}_{\mu\nu} = \partial^{\mu}T^{M}_{\mu\nu} \quad \text{и} \quad \partial^{\nu}T^{E}_{\mu\nu} = \partial^{\nu}T^{M}_{\mu\nu} \quad (2)$$

For a symmetric mechanical energy-momentum tensor, the divergences for each of the indices are equal. Therefore, the right-hand sides of these equations are equal. Divergences for each index exist simultaneously, so they can be written as a sum of divergences by taking the sum of equations (2):

$$(\partial^{\mu}T^{E}_{\mu\nu} + \partial^{\nu}T^{E}_{\mu\nu})/2 = \partial^{\mu}T^{M}_{\mu\nu} \quad (3)$$

Asymmetric tensor $T^{E}_{\mu\nu}$ can be decomposed into an antisymmetric $T^{E}_{[\mu\nu]}$ and symmetric $T^{E}_{(\mu\nu)}$ tensors:

$$T^{E}_{\mu\nu} = \frac{1}{2}\cdot T^{E}_{[\mu\nu]} + \frac{1}{2}\cdot T^{E}_{(\mu\nu)} = \frac{1}{2}\cdot\begin{bmatrix} 0 & i\frac{1}{c}\mathbf{S} - ic\cdot\mathbf{g} \\ -i\frac{1}{c}\mathbf{S} + ic\cdot\mathbf{g} & t_{ik} - t_{ki} \end{bmatrix} + \frac{1}{2}\cdot\begin{bmatrix} 2W & i\frac{1}{c}\mathbf{S} + ic\cdot\mathbf{g} \\ i\frac{1}{c}\mathbf{S} + ic\cdot\mathbf{g} & t_{ik} + t_{ki} \end{bmatrix} \quad (4)$$

It is obvious that the divergences of the antisymmetric tensor $T^{E}_{[\mu\nu]}$ is equal in absolute value and have different signs. Therefore, the sum of its divergences equal to zero. Then the left-hand side of equation (3) will contain only two equal divergences with respect to different indices of the symmetric EMT $T^{E}_{(\mu\nu)}$, and equation (3) can be written in the form:

$$\partial^{\mu}T^{E}_{(\mu\nu)} = \partial^{\mu}T^{M}_{\mu\nu} \quad (5)$$

where, taking into account the coefficient ½ $T^{E}_{(\mu\nu)} = \begin{bmatrix} W & (i\frac{1}{c}\mathbf{S} + ic\cdot\mathbf{g})/2 \\ (i\frac{1}{c}\mathbf{S} + ic\cdot\mathbf{g})/2 & (t_{ik} + t_{ki})/2 \end{bmatrix}$. Now on the right and left sides of this equation there are symmetric EMT, therefore, in this equation the angular momentum is conserved. Then the existing view that an asymmetric EMT violates the law of conservation of angular momentum is incorrect. The reason for the erroneous view is that usually divergence is research only for one of the indices of the energy-momentum tensor. Then in the equation of conservation of the electromagnetic momentum, the Abraham vortex force appears $\mathbf{F}_A = \partial_t \mathbf{g}^M - \partial_t \mathbf{g}^A = \nabla\times(\mathbf{E}\times\mathbf{D} + \mathbf{B}\times\mathbf{H})$ [60]. Then the vortex force of the Abraham leads to the appearance of an uncompensated angular momentum (the effect of Feigele) and to the violation of the angular momentum conservation law. This error leads to the fact that some authors [61 - 67] and others are considering the possibility of extracting this angular momentum from a vacuum. The Abraham force follows from an antisymmetric EMT in the form of its divergence by any of the indices. In full divergence of the asymmetric and antisymmetric EMT, the Abraham force enters with different signs and vanishes. Thus, the complete divergence of the antisymmetric tensor $T^E_{[\mu\nu]}$ proves that the Abraham force is identically equal to zero and does not really exist in nature. Consequently, there is no uncompensated moment of momentum in nature, and it is impossible to extract it from the vacuum.

All complete conservation laws for energy and momentum follow from the symmetric part $T^E_{(\mu\nu)}$ of the asymmetric tensor $T^E_{\mu\nu}$. Therefore, to each asymmetric EMT there corresponds a symmetric EMT associated with it, which can be used for practical calculations.

To asymmetric tensor of Minkowski corresponds (subject to the coefficient ½) the symmetric EMT:

$$W = (\mathbf{E}\cdot\mathbf{D}+\mathbf{H}\cdot\mathbf{B})/2 \qquad \mathbf{S} = (c^2\cdot(\mathbf{D}\times\mathbf{B})+\mathbf{E}\times\mathbf{H})/2$$

$$\mathbf{g} = (\mathbf{D}\times\mathbf{B}+\mathbf{E}\times\mathbf{H}/c^2)/2 \qquad t_{ik} = (E_i D_k + E_k D_i + B_i H_k + B_k H_i)/2 - \delta_{ik}(\mathbf{E}\cdot\mathbf{D}+\mathbf{B}\cdot\mathbf{H})/2$$

To asymmetric EMT obtained in the work [4], corresponds (subject to the coefficient ½) the symmetric EMT:

$$W = \mathbf{E}\cdot\mathbf{D} \qquad \mathbf{S} = (c^2\cdot(\mathbf{D}\times\mathbf{B})+\mathbf{E}\times\mathbf{H})/2$$

$$\mathbf{g} = (\mathbf{D}\times\mathbf{B}+\mathbf{E}\times\mathbf{H}/c^2)/2 \qquad t_{ik} = (E_i D_k + E_k D_i + B_i H_k + B_k H_i)/2 - 3\cdot\delta_{ik}(\mathbf{B}\cdot\mathbf{H}) \qquad (6)$$

This EMT has the remarkable feature that distinguishes it from other known EMT. Its linear invariant (the sum of diagonal components) is the expression $I = \mathbf{E}\cdot\mathbf{D} - \mathbf{B}\cdot\mathbf{H}$. For vacuum, this invariant is the canonical quadratic invariant of the electromagnetic field $I = E^2 - B^2$, which is the density of Lagrangian function of electromagnetic field [68]. This EMT differs from Minkowski's tensor only by diagonal components, but this difference is of principal importance, since it is associated with electromagnetic forces in the medium. From this invariant it follows that the electromagnetic forces in the medium can change sign depending on the ratio of the electric and magnetic characteristics of the medium, associated with the $\mathbf{D}$ and $\mathbf{H}$.

Linear invariant of Minkowski tensor has the form $I^M = \mathbf{E}\cdot\mathbf{D} + \mathbf{B}\cdot\mathbf{H}$. Linear Invariant of tensor Minkowski for the electromagnetic field in vacuum has the form $I^M = E^2 + B^2$. This invariant is quadratic for the electromagnetic field. For an electromagnetic field in a vacuum, there are only two quadratic invariants: $I_1 = E^2 - B^2$ и $I_2 = \mathbf{E}\cdot\mathbf{B}$. From this, it follows that the Minkowski's invariant is invalid. It follows from this that the three-dimensional Minkowski stress tensor describing the electromagnetic forces is also incorrect.

In paper [5] obtained asymmetric tensor for a conducting medium $T^E_{\mu\nu} = \mathbf{A}_\mu \otimes \mathbf{J}_\nu$, where $\mathbf{A}_\mu$ - the electromagnetic potential, $\mathbf{J}_\nu$ - four-dimensional current density. Its components have the form:

$W = \rho \cdot \varphi$ - the density of total electromagnetic energy, where ρ - is the charge density, φ - is the scalar potential of the electromagnetic field;

$\mathbf{g} = \rho \cdot \mathbf{A}$ - the density of electromagnetic momentum, where A - is the vector potential of the electromagnetic field;

$\mathbf{S} = \varphi \cdot \mathbf{J}$ - the flux density of electromagnetic energy, where J - is the current density;

$t_{ik} = -\mathbf{A}_i \otimes \mathbf{J}_k$ - the three-dimensional tensor of flux density of electromagnetic momentum or stress tensor.

This asymmetric tensor also can be decomposed into antisymmetric and symmetric EMT (4). Of the antisymmetric EMT follows an equation of force of Abraham for a conducting medium in the form of its divergence, on any of the indexes $\mathbf{F}_A = \partial_t (\rho \cdot \mathbf{A}) - \frac{1}{c^2} \partial_t (\varphi \cdot \mathbf{J}) = \nabla \times (\mathbf{A} \times \mathbf{J})$. In the complete divergences of asymmetric and antisymmetric tensors, the Abraham force enters with different signs and is equal to zero. To asymmetric tensor for a conducting medium corresponds (subject to the coefficient ½) the symmetric EMT:

$$W = \rho \cdot \varphi \qquad \mathbf{S} = (\varphi \cdot \mathbf{J} + c^2 \cdot \rho \cdot \mathbf{A})/2$$
$$\mathbf{g} = ((\varphi \cdot \mathbf{J})/c^2 + \rho \cdot \mathbf{A})/2 \qquad t_{ik} = (\mathbf{A}_i \otimes \mathbf{J}_k + \mathbf{A}_k \otimes \mathbf{J}_i)/2 \qquad (7)$$

Linear symmetric invariant of this tensor has the form:

$$I = \varphi \cdot \rho - \mathbf{A} \cdot \mathbf{J}$$

This invariant is known in electrodynamics [69] as the density of the Lagrangian of the interaction of electromagnetic field with electrical charges.

### 3. Forms of an electromagnetic momentums for asymmetric tensors

The density of electromagnetic momentum in the dielectric medium in the of Minkowski form has the form $\mathbf{g}^M = \mathbf{D} \times \mathbf{B}$. The density of electromagnetic momentum in the of Abraham form has the form $\mathbf{g}^A = \mathbf{S}/c^2 = \mathbf{E} \times \mathbf{H}/c^2$. Consider the differences of these two forms of electromagnetic momentums. Sommerfeld A. [70] divided the electromagnetic quantities connected with the force and associated with the medium. To the quantities connected with the force, he related the electric field strength $\mathbf{E}$ and the induction of the magnetic field $\mathbf{B}$. To the quantities connected with the medium, he referred to the electric induction $\mathbf{D}$ and the intensity of the magnetic field $\mathbf{H}$. The connections between $\mathbf{E}$ and $\mathbf{D}$, $\mathbf{B}$ and $\mathbf{H}$ are determined by the material equations. For weak electromagnetic fields in isotropic non-ferromagnetic dielectric medium without dispersion is usually take the material equations in the form $\mathbf{D} = \varepsilon \cdot \varepsilon_0 \cdot \mathbf{E}$ и $\mathbf{H} = \mathbf{B}/\mu \cdot \mu_0$, where ε₀ and μ₀ are, respectively, the relative dielectric permittivity and magnetic permeability of the medium. Electric induction $\mathbf{D}$ and the magnetic field strength $\mathbf{H}$, respectively, depend on electric and magnetic characteristics of the medium. Then the density of electromagnetic momentum of Minkowski $\mathbf{g}^M = \mathbf{D} \times \mathbf{B}$, which includes electric induction **D**, describes the portion of the electromagnetic momentum associated with the electrical characteristics of the medium. The density of electromagnetic momentum Abraham $\mathbf{g}^A = \mathbf{E} \times \mathbf{H}/c^2$, which includes the magnetic field strength H describes the part of the electromagnetic momentum associated with the magnetic characteristics of the medium. From this it follows that each of these forms describes only part of the full electromagnetic momentum. Out the asymmetric tensor (6) in the form of the divergences for each of its indices, follow equations of conservation for both forms of momentum density. The full form of the density of electromagnetic momentum in the dielectric medium has the form:

$$\mathbf{g}^E = (\mathbf{g}^M + \mathbf{g}^A)/2 = (\mathbf{D} \times \mathbf{B} + \mathbf{E} \times \mathbf{H}/c^2)/2.$$

Of asimmetryc tensor for conducting medium also follow the two forms of the density of electromagnetic momentum: 1) $\mathbf{g}^E = \rho \cdot \mathbf{A}$ и 2) $\mathbf{g}^E = \varphi \cdot \mathbf{J}/c^2$. The first electromagnetic momentum shape associated with the charges density, and the second shape associated with the speed of their movement. From the decomposition of an asymmetric EMT into a symmetric and antisymmetric EMT, the total shape of the electromagnetic pulse density for a conducting medium follows in the form:

$$\mathbf{g}^E = (\varphi \cdot \mathbf{J}/c^2 + \rho \cdot \mathbf{A})/2$$

This density form of the electromagnetic momentum may find application in the theoretical research of the dynamics of the plasma.

### 4. Full equations of conservation of electromagnetic energy and momentum for an asymmetric tensors

The equations of conservation of electromagnetic energy and momentum are the full four-dimensional divergences of the asymmetric tensor $T^E_{\mu\nu}$: $(\partial^\mu T^E_{\mu\nu} + \partial^\nu T^E_{\mu\nu})/2 = \partial^\mu T^M_{\mu\nu}$ or its symmetric part $\partial^\mu T^E_{(\mu\nu)} = \partial^\mu T^M_{\mu\nu}$. The three-dimensional equations of conservation of density of energy and momentum for dielectric media with losses are of the form:

$$\frac{1}{c}\partial_t (\mathbf{E} \cdot \mathbf{D}) + c \cdot \nabla \cdot (\mathbf{D} \times \mathbf{B} + (\mathbf{E} \times \mathbf{H})/c^2)/2 = \partial_t m + \nabla \cdot \mathbf{p} \qquad (8)$$

$$(\frac{1}{c^2}\partial_t (\mathbf{E} \times \mathbf{H}) + \partial_t (\mathbf{D} \times \mathbf{B}) - \partial_i (E_i D_k + B_i H_k - 3\delta_{ik}(\mathbf{B} \cdot \mathbf{H})) + \partial_k (E_i D_k + B_i H_k - 3\delta_{ik}(\mathbf{B} \cdot \mathbf{H})))/2 = \partial_t \mathbf{p} + \nabla(\mathbf{p} \cdot \mathbf{V})$$

ore $\quad \partial_t(\mathbf{g}^A + \mathbf{g}^M)/2 - (\partial_i(E_i D_k + B_i H_k - 3\delta_{ik}(\mathbf{B} \cdot \mathbf{H})) + \partial_k(E_i D_k + B_i H_k - 3\delta_{ik}(\mathbf{B} \cdot \mathbf{H})))/2 = \partial_t \mathbf{p} + \nabla(\mathbf{p} \cdot \mathbf{V})$, (9)

where m - is the mass density of charges; **p** - is the density of mechanical momentum, **V** - is the velocity of the charges.

If the dielectric medium is described by the canonical constitutive equations with a scalars $\varepsilon$ and $\mu$, the off-diagonal components of the stress tensor $t_{ik}$ equal to zero. Then the electromagnetic forces acting on the medium, are determined only by diagonal components of tensor $t_{ik}$ and equations (8) and (9) can be written as:

$$\frac{\varepsilon \cdot \varepsilon_0}{c} \partial_t E^2 + c \cdot \nabla \cdot (\varepsilon \cdot \varepsilon_0 \mathbf{E} \times \mathbf{B} + (\mathbf{E} \times \mathbf{B})/\mu \cdot \mu_0 \cdot c^2)/2 = \partial_t m + \nabla \cdot \mathbf{p} \qquad (10)$$

$$\partial_t (\mathbf{g}^A + \mathbf{g}^M)/2 - \nabla (\varepsilon \cdot \varepsilon_0 \cdot E^2 - 2 \cdot B^2 / \mu \cdot \mu_0)/2 = \partial_t \mathbf{p} + \nabla (\mathbf{p} \cdot \mathbf{V}) \qquad (11)$$

The full three-dimensional equations of conservation of energy and momentum for a conducting medium (the density of charges and currents) have the following form:

$$\frac{1}{c^2} \partial_t (\varphi \cdot \rho) + (\nabla \cdot (\rho \cdot \mathbf{A}) + \frac{1}{c^2} \nabla \cdot (\varphi \cdot \mathbf{J}))/2 = \partial_t m + \nabla \cdot \mathbf{p} \qquad (12)$$

$$(\frac{1}{c^2} \partial_t (\varphi \cdot \mathbf{J}) + \partial_t (\rho \cdot \mathbf{A}) + \partial_i (\mathbf{A}_i \otimes \mathbf{J}_k) + \partial_k (\mathbf{A}_i \otimes \mathbf{J}_k))/2 = \partial_t \mathbf{p} + \nabla (\mathbf{p} \cdot \mathbf{V}) \qquad (13)$$

Equations (10) and (12) are the complete equations of conservation of energy and equation (9), (11) and (13) are the equations of balance of density of electromagnetic and mechanical forces. Equations (12) and (13) can be used in plasma physics to model the motion of charged particles in an electromagnetic field.

### 5. Conclusion

The widespread opinion that the asymmetric of energy-momentum tensor violates the law of conservation of momentum is incorrect. The full divergence of a tensor of the second rank is equal to the sum of the divergences for each of its indexes. Full asymmetrical divergence of a EMT is the divergence of its symmetric part, i.e. a symmetric EMT. The full divergence of the antisymmetric part is identically equal to zero. The angular momentum for the asymmetric tensor is preserved, since the momentum conservation equations follows from of its symmetric part.

Derived all forms of electromagnetic momentum for dielectric and conducting medium and the complete equations of conservation of energy and momentum. It is shown that the linear invariant of the tensor of Minkowski does not match the quadratic invariant of the electromagnetic field. This indicates the incorrectness of the linear invariant of the of the Minkowski tensor and three-dimensional stress tensor.